\documentclass{IEEEcsmag}

\usepackage[colorlinks,urlcolor=blue,linkcolor=blue,citecolor=blue]{hyperref}

\usepackage{textcomp}
\usepackage{stfloats}
\usepackage{url}
\usepackage{hyperref}
\usepackage{verbatim}
\usepackage{graphicx}

\usepackage{tabularray}
\usepackage[
  backend=biber,
  style=numeric,
  sorting=nyt,
  maxbibnames=6,
  giveninits=true
]{biblatex}
\usepackage{upmath}

\jvol{XX}
\jnum{XX}
\paper{x}
\jmonth{July}
\jname{IEEE Computer Graphics and Applications}

\pubyear{2026}

\begin{document}


\sptitle{Perspective and Overview Paper (POP)}

\title{Towards Insight-Centric Data Spaces: Integrating Visual Analytics and Decentralized Data Ecosystems}

\author{Christina Faßbender}
\affil{Fraunhofer FIT, Sankt Augustin, 53757 , Germany}

\author{Benjamin Greiner}
\affil{Fraunhofer FIT, Sankt Augustin, 53757 , Germany}

\author{Benedikt T. Arnold}
\affil{Fraunhofer FIT, Sankt Augustin, 53757 , Germany}

\author{Georgios Toubekis}
\affil{Fraunhofer FIT, Sankt Augustin, 53757 , Germany}

\author{Christoph Lange}
\affil{Fraunhofer FIT, Sankt Augustin, 53757 , Germany}

\markboth{Perspective and Overview Paper (POP)}{Perspective and Overview Paper (POP)}



\begin{abstract}
Visual analytics and dataspaces address complementary challenges: visual analytics supports interactive exploration and human-centered reasoning, while dataspaces enable decentralized, sovereign data sharing. Despite their potential synergies, both fields have largely evolved independently. We examine their intersection, positioning visual analytics as a cognitive layer that makes data availability, quality, provenance, and exchange processes interpretable, while dataspaces introduce new requirements for distributed, heterogeneous, and governed analysis. A semi-synthetic cultural dataspace scenario illustrates opportunities for exploring exchange patterns and metadata quality across federated participants. We identify open challenges and envision insight-centric dataspaces that integrate data management and visual analysis into a unified, human-centered environment.
\end{abstract}

\maketitle

\chapteri{V}isual analytics has been established as a powerful approach for understanding and communicating complex and heterogeneous data. By combining automated analysis with interactive visual representations, it enables users to explore data, uncover patterns, and support decision-making processes. However, the effectiveness of visual analytics strongly depends on how data is stored, managed, and prepared prior to analysis.

Consequently, data management has evolved into a dedicated discipline that addresses the storage, integration, and access of data across different environments~\cite{DAMA2017}. Traditional approaches, such as centralized databases or cloud infrastructures, assume that data can be consolidated and controlled within a single system. In contrast, modern data ecosystems are increasingly decentralized, heterogeneous, and governed by access constraints, making such assumptions less realistic.

In this context, dataspaces have emerged as a promising paradigm for decentralized data management. They enable sovereign data exchange across organizational boundaries while allowing data owners to retain control over their data. By supporting controlled data sharing through negotiated usage policies, dataspaces facilitate collaboration in distributed data ecosystems without requiring full upfront integration~\cite{JarkeOttoEtAl2019}. 

Although the concept of dataspaces has been introduced already some time ago~\cite{halevy2006principles}, and design principles as well as architectural frameworks are increasingly well understood and standardized~\cite{CurryScerriEtAl2022}, many challenges remain in their practical adoption. In particular, current dataspace implementations focus primarily on data exchange, governance, and interoperability, but provide limited support for understanding, analyzing, and assessing the data within these environments.

This limitation highlights a fundamental gap: while dataspaces enable access to distributed data, they do not inherently support sensemaking. At the same time, visual analytics typically assumes that data is already integrated and readily available. As a result, both fields address complementary aspects of the data pipeline, but remain insufficiently connected.

In this work, we argue that visual analytics and dataspaces form a highly promising intersection. Visual analytics can act as an interactive and human-centered layer for dataspaces, enabling users to explore data availability, quality, provenance, and exchange processes. Conversely, dataspaces introduce new challenges and opportunities for visual analytics by providing decentralized, heterogeneous, and governed data environments.

To illustrate this interplay, we present a use case from the cultural domain, the \textit{Culture Dataspace (Datenraum Kultur)}, and demonstrate how visual analytics can support its analysis and operation. \textit{Datenraum Kultur} is one of the lighthouse projects of the German government's National Digital Strategy~\cite{acatec2022}, aiming to enable sovereign data exchange across the cultural and creative sector~\cite{Jarke2023}. 

Building on this example, we first analyze how visual analytics can enhance dataspaces, and subsequently how dataspaces can reshape visual analytics systems. Finally, we identify open research challenges and outline future directions for strengthening this interdisciplinary field.

Therefore, this paper contributes:
\begin{itemize}
    \item A short introduction to dataspaces
    \item A structured description of the intersection between visual analytics and dataspaces
    \item A real-world inspired use case demonstrating this intersection
    \item Open research problems and future directions
\end{itemize}

\section{Basics of Dataspaces}

Dataspaces support collaboration across distributed, heterogeneous
data sources through incremental integration rather than requiring
comprehensive harmonization upfront
\cite{otto2019designing,theissen2024towards, HalevyFM2006}.
Contemporary sovereign dataspaces complement this approach with
participant identification, asset discovery, and agreements governing
access and use \cite{JarkeOttoEtAl2019}. Their distinguishing concern
is not the absence of schemas or central services, but cooperation
across organizational boundaries while preserving participants'
control over their assets.

\subsection{Infrastructure and Workflow}

\begin{figure}[t]
    \centering
    \includegraphics[width=\linewidth]{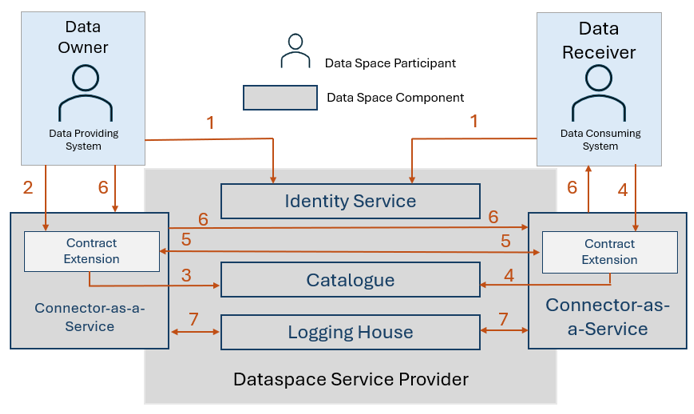}
    \caption{Dataspace infrastructure and exchange workflow.
    Catalogues, identity services, and contract management support
    discovery and agreement negotiation; connectors enable exchange,
    while logging can support subsequent monitoring.}
    \label{fig:dataspace-architecture}
\end{figure}

Figure~\ref{fig:dataspace-architecture} summarizes the infrastructure.
Providers describe assets in catalogues, consumers discover relevant
offerings, identity services establish participant identities, and
connectors support agreement negotiation and authorized exchange.
Logging can support accountability where implemented and accessible.
A typical workflow comprises onboarding, discovery, authentication,
agreement negotiation, and transfer or access. These mechanisms
coordinate sharing; they do not automatically guarantee semantic
compatibility, data quality, or compliance with every usage condition.

\subsection{Practical Applications and Opportunities for VA}

Existing dataspace initiatives illustrate the potential of governed,
cross-organizational data sharing and emerging opportunities for
visual analytics (VA).

\paragraph{Automotive supply chains: Catena-X.}
Catena-X supports traceability, quality management, circularity,
and capacity planning across automotive supply chains.
\footnote{\url{https://catena-x.net/overview-use-case-cluster/}}
VA could reveal supplier dependencies, trace faulty components,
and compare quality observations across organizations, while
distinguishing missing information from confirmed findings.

\paragraph{Transport and road safety: Mobility Data Space.}
The Mobility Data Space enables applications such as winter road
maintenance and parking occupancy monitoring.
\footnote{\url{https://mobility-dataspace.eu/use-cases}}
VA could reveal geographic risk patterns, temporal trends,
and discrepancies between observations and forecasts, while
making data freshness, coverage, and resolution transparent.

\paragraph{Health research: European Health Data Space.}
The European Health Data Space provides a framework for governed
secondary use of health data across institutions and countries.
\footnote{\url{https://health.ec.europa.eu/ehealth-digital-health-and-care/reuse-health-data_en}}
Within authorized environments, VA could support cohort exploration,
population comparisons, and identification of missingness or bias,
while respecting privacy and access restrictions.

\paragraph{Culture: distributed collections and services.}
The Culture Dataspace connects cultural institutions and services
across heterogeneous collections \cite{drk}.
VA could support collection discovery, metadata quality assessment,
and identification of enrichment opportunities while accounting
for usage rights. Our later scenario illustrates these possibilities
using semi-synthetic data rather than actual institutional activity.

\subsection{Relevance to Visual Analytics}

Across these domains, dataspaces matter because useful evidence
is distributed among actors with different responsibilities,
technical systems, and disclosure constraints. Infrastructure can
make collaboration possible, but analysts still need to understand
which data are available, comparable, trustworthy, and usable for
their purpose.

This creates a two-way relationship with VA. Dataspaces can provide
governed access to distributed evidence; VA can help users inspect
its coverage, quality, provenance, and integration choices.
Conversely, dataspace settings require visual interfaces to make
access restrictions, unavailable evidence, and uncertainty explicit.
Their value therefore extends beyond transporting data to supporting
informed reasoning across organizational boundaries.

\section{Why should we care about this intersection?}

Visual analytics and dataspaces address complementary challenges in data-driven decision making. Visual analytics supports interactive exploration, sensemaking, and decision support \cite{Keim2008}, while dataspaces enable decentralized, sovereign, and incremental integration of heterogeneous data across organizational boundaries \cite{CurryScerriEtAl2022,JarkeOttoEtAl2019}. Their intersection offers significant potential but remains insufficiently explored.

\textbf{Bridging data integration and analysis.}
Visual analytics often assumes centrally accessible, pre-integrated data, whereas dataspaces follow a \emph{pay-as-you-go} integration model. Interactive data preparation tools such as Wrangler \cite{kandel2011wrangler} and knowledge-enhanced systems such as CAVA \cite{9222249} demonstrate how integration and analysis can be coupled. However, they do not explicitly address decentralized ownership, sovereignty, or contract-based access.

\textbf{Handling heterogeneity and scale.}
Integrating data management with visualization has long been recognized as essential for analyzing large, heterogeneous datasets \cite{gray2005scientific,fekete2012managing}. Systems such as Tioga and VisTrails illustrate this connection \cite{silva2011tioga}. Dataspaces extend these concepts to distributed, independently governed data ecosystems, introducing additional requirements for discovery, integration, and interactive analysis.

\textbf{Semantic and collaborative exploration.}
Semantic technologies support interoperability across heterogeneous sources, while visual interfaces help users explore complex ontologies and relationships \cite{gutierrez2018interactive}. Combining these capabilities with dataspace infrastructures could enable collaborative visual analytics across organizational boundaries while respecting ownership and usage constraints.

\textbf{From isolated systems to data ecosystems.}
Existing research highlights the growing importance of integrated data management and analysis in complex data environments \cite{sivarajah2017critical,al2017survey}. Dataspaces provide the infrastructure for decentralized exchange, while visual analytics can make distributed data, relationships, and governance constraints accessible and understandable to users.

\textbf{The research gap.}
Despite these complementary strengths, integrated approaches remain limited. Visual analytics research primarily addresses data exploration and analytical workflows, whereas dataspace research emphasizes infrastructure, interoperability, governance, and exchange. How visual analytics can operate within sovereign data ecosystems, and how dataspace characteristics reshape visualization requirements, remains insufficiently investigated \cite{zhang2012visual}.

Addressing this gap is essential for enabling interactive, collaborative, and trustworthy analysis across organizational boundaries. Visual analytics can provide the human-centered interface to dataspaces, while dataspaces offer the decentralized infrastructure for accessing and integrating distributed data. Their intersection thus opens new research opportunities for visualization and data management alike.

\section{The intersection of visual analytics and dataspaces}

\subsection{Visual Analytics for Dataspaces}

We frame VA opportunities around concrete cross-organizational
tasks rather than generic visualization capabilities. For each
opportunity, we identify the user, their goal, the dataspace-specific
difficulty, and the proposed analytical support. These are design
propositions requiring evaluation, not demonstrated benefits.
In particular, asset discovery should not be equated with ordinary
data exploration, and real-time support is not an inherent
dataspace capability.

\paragraph{Enhanced discovery: assessing distributed offerings.}
\emph{User and task:} A cultural platform curator preparing a regional
exhibition must select complementary assets from museums, archives,
and libraries, covering relevant themes, places, and periods under
suitable usage conditions.
\emph{Dataspace-specific difficulty:} Institutions independently
maintain their descriptions and control access. Individually relevant
search results may use incompatible vocabularies, leave coverage
gaps, or have conditions that prevent their intended joint reuse.
\emph{VA opportunity:} Coordinated thematic, spatial, and temporal
views could support comparison of candidate collections alongside
metadata coverage and declared usage conditions. The curator could
identify, for example, an offering that fills a geographic gap but
lacks temporal descriptions or republication rights. The proposed
contribution is assessing combinations of distributed offerings,
not replacing keyword search with a graphical interface. Whether
this improves selection over conventional catalog search remains
a research question; visible metadata does not imply authorized
access to the underlying assets.

\paragraph{Integration and schema alignment.}
\emph{User and task:} A data steward integrating regional event
platforms must align venue identifiers, event categories, dates,
and accessibility descriptions.
\emph{Dataspace-specific difficulty:} Providers retain independent
schemas and update practices; no participant can unilaterally
harmonize all sources.
\emph{VA opportunity:} Interactive mapping views could expose
candidate correspondences, conflicting meanings, and the effects
of alternative transformations on authorized examples.
Schema-mapping research provides a starting point \cite{Coscia2024},
but mappings still require domain validation. The relevant benefit
would be better-informed integration decisions across providers,
not automatic interoperability.

\paragraph{Provenance and exchange investigation.}
\emph{User and task:} An automotive quality investigator tracing
a defective component must identify contributing suppliers and
potentially affected downstream products.
\emph{Dataspace-specific difficulty:} Evidence spans organizational
boundaries, with different identifiers, disclosure permissions,
and independently maintained histories.
\emph{VA opportunity:} Linked views of authorized asset identities,
exchange events, and transformation records could support dependency
tracing while marking missing evidence
\cite{7120975,Yazici2023}. Exchange logs alone do not establish
record-level lineage. VA can support investigation only where
appropriate histories exist and are accessible; it cannot infer
a complete chain from an incomplete visible graph.

\paragraph{Metadata quality assessment and curation.}
\emph{User and task:} A federation metadata coordinator must identify
recurring description problems and prioritize support for providers.
\emph{Dataspace-specific difficulty:} Independently maintained
catalogs contribute to shared discovery, while correction
responsibilities remain distributed.
\emph{VA opportunity:} Field-level comparisons, quality distributions,
and record-level drill-down could reveal whether poor discovery
coverage arises from missing descriptions, incompatible terminology,
or uneven provider practices. Quality and uncertainty visualization
offer relevant foundations
\cite{2017-1,maack2023uncertainty,gillmann2023taxonomy}.
Indicators must be task-specific: metadata completeness does not
establish correctness or fitness for reuse. Our semi-synthetic
scenario illustrates this inspection workflow without identifying
actual institutional deficits.

\paragraph{Federated exchange and service monitoring.}
\emph{User and task:} A dataspace operations coordinator investigating
delayed deliveries must distinguish provider unavailability,
connector failures, access rejection, and downstream processing
delays.
\emph{Dataspace-specific difficulty:} A delivery crosses independently
operated services whose telemetry and monitoring permissions differ.
\emph{VA opportunity:} Correlated request timelines, failure categories,
and latency distributions could help locate interruptions and
inspect unusual exchange patterns. This requires comparable,
authorized telemetry and does not grant centralized control over
participants. Generic anomaly detection and resource monitoring
are therefore included here only insofar as they support diagnosing
cross-participant workflows.

\paragraph{Real-time analysis and decision support:
a conditional research hypothesis.}
\emph{User and task:} A road-maintenance dispatcher combines
vehicle-derived road conditions, weather, and operational data
to prioritize gritting.
\emph{Dataspace-specific difficulty:} Independent providers
impose varying access conditions, freshness, coverage, and reliability.
\emph{VA opportunity:} Given timely, authorized feeds, linked maps
and timelines could highlight hazards, outdated observations,
coverage gaps, and conflicting evidence to support decisions.

However, dataspaces do not guarantee streaming access, low latency,
or availability. This remains a research hypothesis requiring
evaluation against existing workflows under realistic constraints.

More broadly, these opportunities depend on governed evidence,
suitable metadata, and user evaluation. Graphical querying,
alerting, and dashboards are supporting mechanisms rather than
independent contributions.

\subsection{Dataspaces for Visual Analytics}

Dataspaces may support VA by enabling governed access to evidence
held by independent organizations. However, they do not inherently
provide harmonized semantics, scalable computation, real-time feeds,
or collaborative interfaces. We therefore connect each opportunity
to a concrete analytical task and distinguish enabling infrastructure
from capabilities that require additional implementation and evaluation.

\paragraph{Semantic alignment across providers.}
\emph{User and task:} A transport analyst comparing service coverage
across municipalities must reconcile different definitions of stops,
routes, service periods, and accessibility.
\emph{Dataspace-specific difficulty:} These descriptions originate
from independently maintained systems; similarly named attributes
may represent different concepts.
\emph{Dataspace contribution:} Shared vocabularies, published schemas,
and documented mappings, where adopted, could provide VA systems
with explicit semantic relationships. Analysts could then inspect
which comparisons are supported and where harmonization remains
incomplete. Ontology-oriented VA offers a methodological starting
point \cite{8440124}. Dataspaces provide a setting for negotiating
alignment, not a guarantee of semantic equivalence.

\paragraph{Governed enrichment of analytical evidence.}
\emph{User and task:} A cultural collection analyst investigating
geographic gaps must determine whether they reflect limited holdings,
incomplete descriptions, or missing contributions from other institutions.
\emph{Dataspace-specific difficulty:} Relevant collection descriptions
and contextual information belong to different organizations and
may have distinct access conditions.
\emph{Dataspace contribution:} Catalog discovery and authorized access
could allow the analyst to introduce additional evidence during
exploration rather than rely solely on a fixed initial dataset.
VA could expose how enrichment changes coverage and interpretation,
while preserving distinctions between sources and their conditions
of use. Dataspace-enabled analytics motivates this direction
\cite{gillmann_dataspace-enabled_2024}; reproducibility additionally
requires recording asset versions, transformations, and access context.

\paragraph{Analysis without unrestricted data centralization.}
\emph{User and task:} A health researcher comparing outcomes across
institutions needs suitable population summaries while respecting
restrictions on patient-level information.
\emph{Dataspace-specific difficulty:} Institutions control sensitive
records under different legal and organizational requirements.
\emph{Dataspace contribution:} Governed access arrangements could
support approved aggregates or analysis within authorized environments.
These outputs could inform VA without requiring unrestricted
transfer of all underlying records. However, federated computation,
privacy protection, and disclosure control require additional
mechanisms. Whether analysts can reason effectively from restricted
or uneven summaries is a research question; this opportunity is
not evidence that dataspaces automatically improve scalability
or computational performance.

\paragraph{Cross-organizational analytical collaboration.}
\emph{User and task:} A manufacturer and its suppliers jointly
investigating component failures must compare observations and
discuss competing explanations.
\emph{Dataspace-specific difficulty:} Participants have different
permissions, responsibilities, and access to supporting evidence.
\emph{Dataspace contribution:} Participant identities and negotiated
sharing arrangements could establish which evidence may enter
a joint analysis. A collaborative VA layer could associate findings
with accessible sources and make differences in participants'
visible evidence explicit. Shared annotations, synchronized views,
and permissions for derived results require separate implementation.
Improved collective reasoning is therefore a hypothesis to evaluate,
not an automatic consequence of data sharing.

\paragraph{Continuously updated analysis:
a conditional research hypothesis.}
\emph{User and task:} A municipal mobility analyst monitoring parking
pressure needs observations and forecasts from multiple operators.
\emph{Dataspace-specific difficulty:} Feeds have independent owners,
access agreements, update schedules, and service guarantees.
\emph{Dataspace contribution:} Where agreements and interfaces permit,
a dataspace could make these feeds available to a VA workflow.
The interface could distinguish current observations, stale values,
forecasts, and unavailable sources. Neither continuous delivery
nor interactive latency is guaranteed by the dataspace itself.
The research question is whether governed multi-provider updates
improve situational understanding sufficiently to justify their
integration and operational complexity.

Taken together, these opportunities position dataspaces as potential
enablers of cross-organizational VA, not substitutes for analytical
infrastructure. We therefore do not retain scalability, multimodal
support, or flexibility as independent benefits: their relevance
depends on a specific task, available assets, and implemented services.
The central hypothesis is that governed access to complementary
evidence can improve analytical reasoning, provided that semantic
differences, access restrictions, provenance, and uncertainty remain
visible throughout the workflow.

\section{The Culture Dataspace Scenario: Monitoring Semi-Synthetic Data}
\label{sec:culture-scenario}

The following scenario illustrates how visual analytics could support metadata quality assessment, exchange observability, and resource monitoring within a federated cultural data ecosystem. It is a vision-oriented design demonstrator, not an operational dashboard or an empirical evaluation of the current Culture Dataspace. In particular, displayed source--destination relations suggest opportunities for exchange monitoring; they do not establish record-level provenance, transformation lineage, or policy compliance.

The discussion follows the opportunity categories introduced in \emph{The Intersection of Visual Analytics and Dataspaces}. The exchange overview and Bautzen detail illustrate \emph{federated exchange and service monitoring}, limited here to exchange structure; the stacked quality view illustrates \emph{metadata quality assessment and curation}. Other categories are explicitly identified below as requirements or future extensions rather than demonstrated capabilities.

\subsection{Cultural Context and Analytical Motivation}

Initiatives such as the \emph{common European dataspace for cultural heritage}\footnote{\url{https://www.dataspace-culturalheritage.eu/en}} seek to facilitate cross-organizational and cross-sectoral exchange through shared technical infrastructure and governance frameworks. In Germany, the \textit{Culture Dataspace} pursues trusted, decentralized exchange while allowing participating institutions to retain sovereign control over their assets. Its emphasis on business-to-business exchange aims to enable reuse and value-added services, ultimately supporting offerings for end users despite substantial technical, semantic, legal, and organizational complexity \cite{drk}.

The project pursues efficiency gains through digital networking, support for smaller cultural institutions, and data-driven value creation. Connected cultural platforms, smart museum services, smart theater services, and smart music school services provide settings for eliciting requirements and exploring distributed value creation \cite{drk,ToubekisDecker2025}. These settings encompass content, system-generated data, personal data, and metadata, with heterogeneous access and usage requirements rather than a single distinction between open and restricted assets. Linked Data-based information modeling supports discovery and interoperability \cite{DRK-IM}, while connector infrastructure and machine-readable policies address controlled exchange \cite{mustafa_etal_2025}.

Our scenario focuses on the metadata layer because discoverability and reuse depend on sufficiently informative descriptions. It does not assess the intrinsic quality of cultural objects or simulate the enforcement of access policies. Instead, it explores how a future monitoring interface might expose participation asymmetries and metadata curation needs without revealing sensitive institutional traces.

\subsection{Construction of the Semi-Synthetic Scenario}
\label{sec:semi-synthetic-method}

\paragraph{Sources and preprocessing.}
The visualization snapshot was generated on 15 April 2026. Its geographic foundation is the BKG VG250 district layer (\texttt{vg250:vg250\_krs}, reference date 1 January 2025), containing 433 geometry features for 400 distinct district identifiers. Features were retained rather than dissolved by identifier; approximate polygon areas and area-weighted centroids were calculated for placement and allocation. Consequently, geometry features must not be interpreted as distinct administrative districts.

The metadata basis is a cached sample of 500 GovData CKAN dataset records, requested through \texttt{package\_search} with \texttt{rows=500} and \texttt{sort=metadata\_modified+desc}.\footnote{GovData provides federated access to German public-sector datasets: \url{https://www.govdata.de/}.} All 500 records originate from Open Data Brandenburg, with modification timestamps on 15 April 2026; the cache does not record its retrieval time. This is a convenience sample of administrative metadata, neither representative of GovData nor filtered for cultural content. CKAN extra fields were mapped by key, and null, blank, or empty values were treated as missing. Dataset contents were not downloaded.

\paragraph{Regional redistribution and synthetic exchange.}
Stored state-level GovData search counts provide inventory signals; the snapshot uses fallback counts, including a national catalog total of 151,084, rather than a successful live retrieval at generation time. Counts represent catalog records, not bytes, and state search results need not partition the national catalog. Within each state, its count was redistributed across geometry features in proportion to $A_i^{0.72}u_i$, where $A_i$ is approximate area and $u_i$ is 2.2 for city districts, at least 2.6 for Berlin and Hamburg, and 1 otherwise. Allocations were independently rounded with a minimum of one record per feature, so exact conservation is not guaranteed. These inventories are synthetic regional assignments, not observed local holdings.

The generator deterministically selected 860 directed cross-state relations. Candidate scores combine normalized source and destination inventories, distance decay, urban and regional multipliers, and identifier-derived provider/consumer biases. Per-node limits of 3, 6, or 10 relations and directed state-pair limits of 10 or 14 restrict density. These rules deliberately produce unequal contribution and consumption patterns; they are scenario assumptions, not estimated cultural exchange behavior. Relation magnitudes are score-scaled synthetic traffic units, not measured transfers or counts of distinct shared records. A relation has one source and one destination. Virtual state hubs bundle its rendered path without introducing intermediate transactions or asserting actual mediation infrastructure.

\paragraph{Metadata completeness proxy.}
For each sampled record, the raw score is
\[
 r=\max\left(0,\sum_k w_k c_k-P\right).
\]
Weights for title, description, license, tags, publisher, contact name, contact point, issued date, modified date, language, theme, and spatial coverage are respectively $(.10,.22,.14,.10,.10,.09,.09,.05,.05,.02,.02,.02)$. Components are binary presence indicators except description and tags. Trimmed description lengths below 40, 40--119, 120--279, and at least 280 characters receive scores $(0,.30,.65,1)$; tag counts of 0, 1--2, 3--4, and at least 5 receive $(0,.35,.70,1)$.

License presence accepts \texttt{license\_title} or \texttt{license\_id}; publisher presence accepts the organization title or \texttt{publisher\_name}. Contact name accepts \texttt{maintainer} or \texttt{contact\_name}, and contact point accepts \texttt{maintainer\_email}, \texttt{contact\_email}, or \texttt{contact\_url}. Issued date, language, theme, and spatial coverage use their extra fields; modified date accepts \texttt{modified} or CKAN's \texttt{metadata\_modified}.

The additive penalty $P$ comprises .12 for missing license; .16 for descriptions below 40 characters or .06 for 40--119 characters; .10 when both contact components are absent; .05 when both spatial coverage and theme are absent; and .08 for no tags or .05 for 1--2 tags. Raw scores are rounded to four decimals. This heuristic measures metadata availability and descriptive richness, not correctness, schema conformance, or underlying data quality.

\paragraph{Quality redistribution and normalization.}
Observed raw scores range from .506 to .740. For visual contrast, they are transformed to
\[
 q=\left(\frac{r-.506}{.740-.506}\right)^{.82}.
\]
Thus, $q=1$ denotes the best sampled metadata, not absolute completeness. Scores are deterministically permuted and reused across regions without inferring their actual ownership. For inventory $N_i$, each feature receives $\operatorname{clip}_{[6,36]}(\operatorname{round}(4+\sqrt{N_i}/2.2))$ sampled scores. Their frequencies, rounded to the nearest .05 quality level, determine stacked segment shares; their arithmetic mean determines the regional ranking. These assignments create synthetic regional profiles from real metadata characteristics, not empirical assessments of cultural institutions.

\subsection{VA for Dataspaces: Federated Exchange and Service Monitoring}

\begin{figure}[t]
    \centering
    \includegraphics[width=\linewidth]{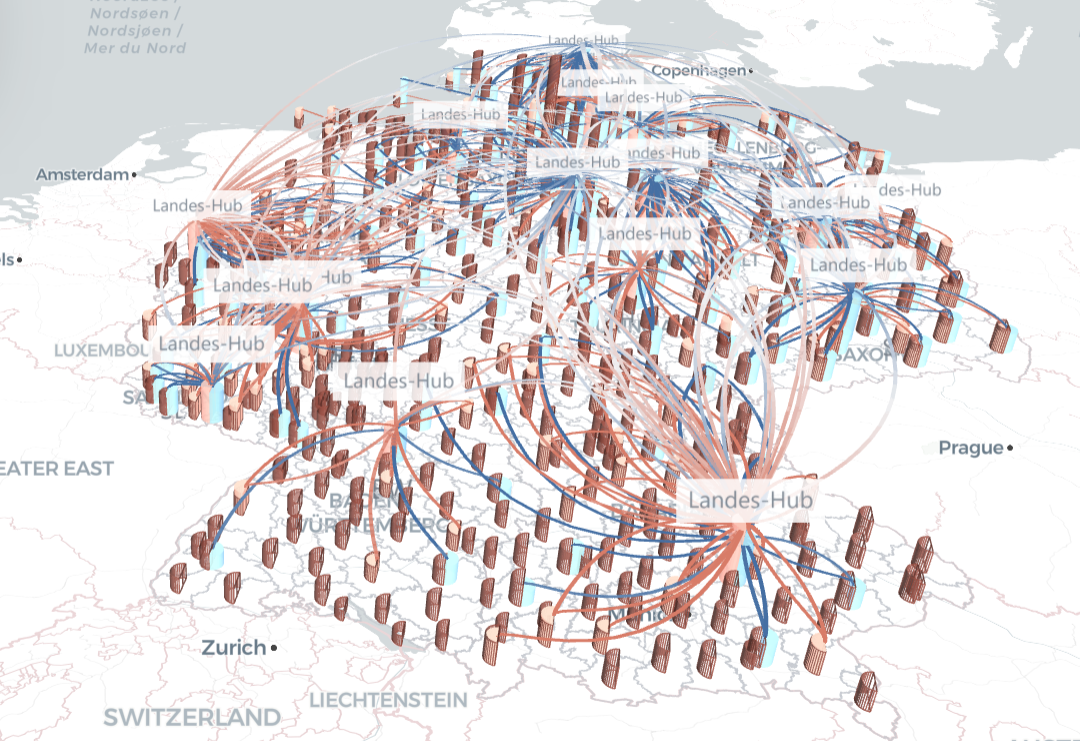}
    \caption{National exchange overview (semi-synthetic, illustrative). White geography and gray boundaries provide spatial context; state abbreviations indicate locations. Paired salmon and blue half-cylinders encode outgoing and incoming synthetic traffic at district centroids and aggregated virtual state hubs. Nonzero heights follow $h=h_{\min}+(h_{\max}-h_{\min})\sqrt{v/M}$, with separate maxima and bounds of 7,000--82,000 (districts) and 12,000--52,000 (hubs) rendering meters; zero traffic is omitted. Heights are exaggerated and not directly comparable across levels. Dark-salmon wireframes indicate catalog-record inventory, not utilization. Curves connect districts via state hubs: salmon source legs, salmon--gray--blue inter-hub gradients, and blue destination legs indicate direction. Width $2300+1.7v$ encodes synthetic relation weight, not bytes or record counts. Cylinder diameter and curve height are layout choices, not data measures.}
    \label{fig:overview}
\end{figure}

\begin{figure}[t]
    \centering
    \includegraphics[width=\linewidth]{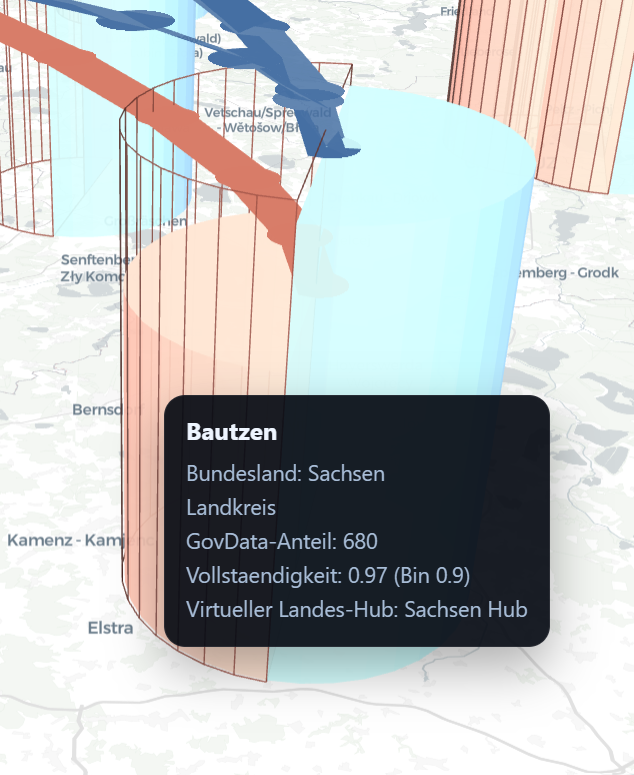}
    \caption{ Bautzen detail (semi-synthetic, illustrative). Co-centered salmon (west) and blue (east) half-cylinders encode outgoing and incoming traffic; the taller extends above the shorter. Zero values have zero height; nonzero district heights follow $h=7000+75000\sqrt{v/M}$ rendering meters, with $M$ the maximum district traffic or inventory value. Dark-salmon wireframes encode catalog-record inventories using the same scale, but heights do not represent sharing fractions. Diameter is unrelated to magnitude. Curves connect directional anchors via virtual state hubs: salmon indicates source legs, blue destination legs, and salmon--gray--blue gradients inter-hub connections. Width $2300+1.7v$ encodes synthetic relation weight, not bytes or dataset counts. Curvature and exaggerated heights improve visibility, not geographic accuracy. White geography and gray boundaries provide context; values do not represent actual Bautzen institutions.}
    \label{fig:bautzen}
\end{figure}

\paragraph{User, task, and scenario element.}
A federation coordinator uses this view to identify provider/consumer imbalances and exchange patterns requiring investigation across independently controlled participants. Figures~\ref{fig:overview} and \ref{fig:bautzen} illustrate linked national and local perspectives. The prototype visualizes synthetic exchanges, not operational telemetry, and cannot diagnose latency, connector failures, or access rejection.

Figure~\ref{fig:overview} shows a spatial hierarchy of local nodes and virtual state hubs, highlighting uneven participation. Cross-state exchanges follow source--source hub--destination hub--destination routes. Hubs serve as visual aggregation points, not actual coordination entities, illustrating how large federations could be explored without relying on flat graphs.

The exchange view supports three related questions: which regions act predominantly as providers or consumers, where contribution and consumption are asymmetric, and how those patterns aggregate at state level. Figure~\ref{fig:bautzen} explains the local encoding through a zoom into Bautzen. Co-located half-cylinders distinguish outgoing exchange in light salmon from incoming exchange in light blue. Where their heights differ, the taller half extends above the shorter one, exposing directional imbalance within a compact footprint. Connections meet the corresponding directional geometry, and the same principle applies to state hubs.

The dark-salmon wireframe supplies an additional inventory-based reference for potential outgoing contribution. It should be read as a scenario comparison cue, not a measured utilization percentage: inventory counts and synthetic traffic magnitudes have different meanings, and nonlinear height scaling prevents direct interpretation of height ratios as exact fractions. Likewise, the Bautzen detail explains the design rather than making a claim about Bautzen's actual cultural exchange activity.

\subsection{VA for Dataspaces: Metadata Quality Assessment and Curation}

\begin{figure}[t]
    \centering
    \includegraphics[width=\linewidth]{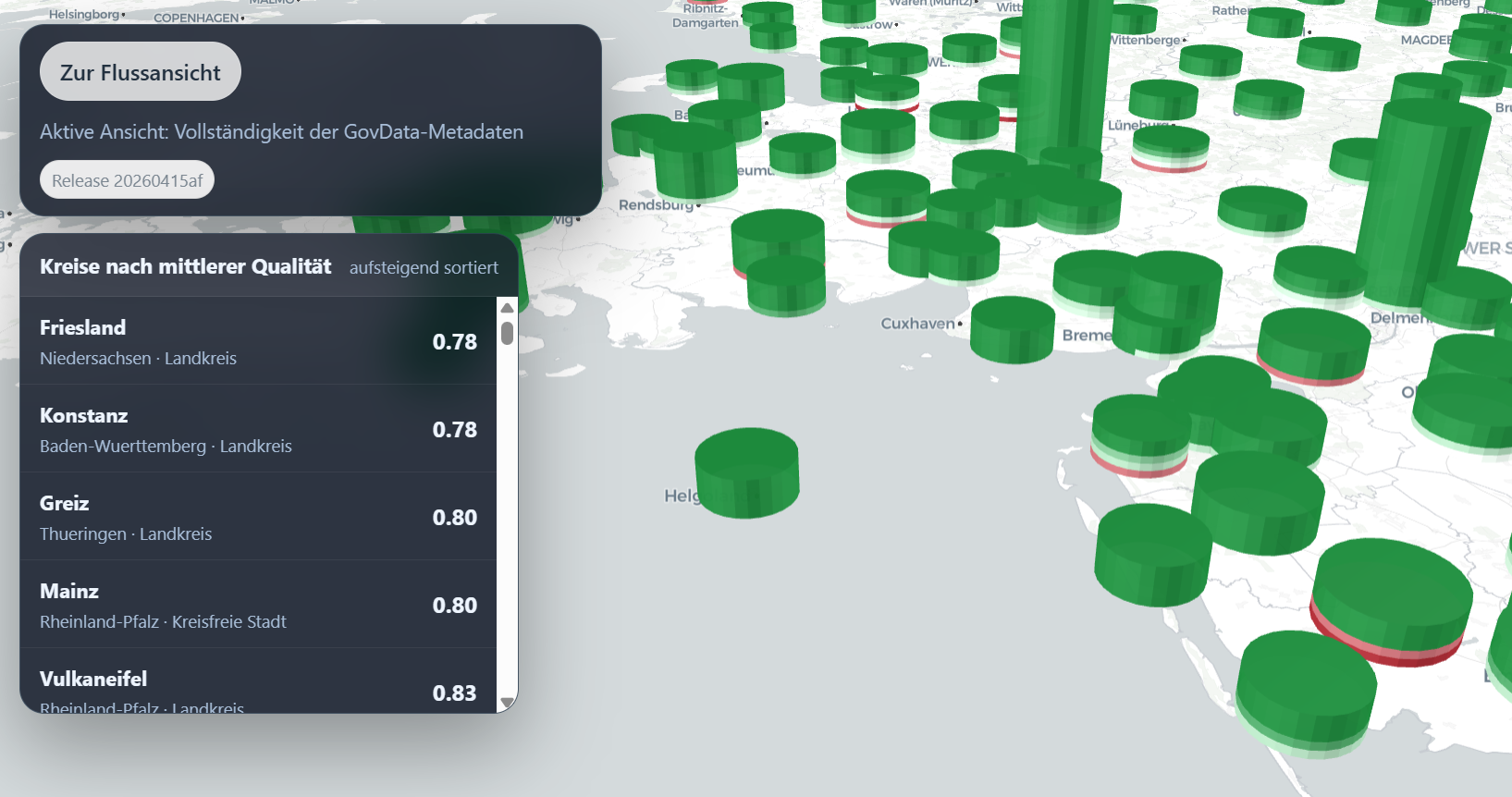}
    \caption{Metadata quality view (semi-synthetic, illustrative). Cylinders encode assigned regional inventories in catalog records, not bytes; display height follows $\max(6000,\operatorname(0.24E_i))$ rendering meters, not a linear count scale or physical altitude. Stacked segments show frequencies of 6--36 sample scores, ordered from low to high. Dimensionless quality scores follow $q=((r-.506)/.234)^0.82$, rounded to .05 increments. Colors range from dark red ($q=0$) through near-white ($.5$) to dark green ($1$), indicating relative sample scores, not validated quality classes or metadata completeness. Segment heights approximate frequency shares, subject to a 1,200-meter minimum; diameter is unrelated to quality. The sidebar ranks regions by ascending mean $q$. White geography and gray boundaries provide context without quality-coded fills. Neither colors nor rankings indicate actual regional metadata deficiencies.}
    \label{fig:quality}
\end{figure}

\paragraph{User, task, and scenario element.}
A federation metadata coordinator would compare provider profiles to prioritize curation support. The dataspace-specific challenge is that shared discoverability depends on descriptions maintained by independent institutions, with distributed responsibility for correction. Figure~\ref{fig:quality} illustrates distribution-aware comparison: ranking helps locate a low-scoring profile, while stacked segments show whether its assigned records are uniformly weak or internally heterogeneous. The demonstrator uses regional proxies rather than actual provider identities and does not implement record-level remediation.

Figure~\ref{fig:quality} switches the analytical focus from exchange to metadata readiness. One stacked cylinder replaces the directional pair at each location. Its height is derived from the assigned inventory, while segment shares represent the local synthetic sample's quality distribution. Low relative scores appear in dark red at the bottom, intermediate scores in lighter tones, and high relative scores in dark green at the top. A sidebar orders regions by their mean score, allowing a low-ranked region to be selected and inspected spatially.

The selected low-ranked region illustrates why distributions matter alongside averages. A node can combine comparatively well-described records with poorly described ones; stacked segments reveal this composition rather than presenting a uniform regional quality label. However, green segments indicate the upper end of this specific sample's normalized score range, not fully complete records. The regional differences are generated assignments, and the figure therefore demonstrates a possible curation workflow rather than identifying actual local deficits. In an operational deployment, such a workflow could help prioritize metadata support for smaller institutions and investigate barriers to discovery and reuse.

\subsection{VA for Dataspaces: Opportunities Beyond the Demonstrator}

\paragraph{Enhanced discovery: assessing distributed offerings.}
The quality view illustrates one prerequisite for discovery, not the asset-selection task itself. Supporting a curator selecting complementary exhibition assets would additionally require searchable asset descriptions, thematic and temporal coverage, and declared usage conditions. The present map cannot establish that offerings are suitable for joint reuse; whether such coordinated views improve selection remains a research hypothesis.

\paragraph{Integration and schema alignment.}
The Culture Dataspace's heterogeneous descriptions motivate a data steward's mapping task, but the prototype does not compare schemas or validate transformations. Extending it would require explicit vocabularies, candidate mappings, and authorized examples through which the steward could inspect cross-provider semantic conflicts.

\paragraph{Provenance and exchange investigation.}
Source--destination curves illustrate exchange relationships only. An investigator tracing a cultural asset's reuse would need persistent asset identities, versions, and transformation histories across institutions. Routing through virtual hubs neither supplies that lineage nor indicates intermediate processing.

\paragraph{Real-time analysis and decision support.}
The snapshot is static and illustrates no time-sensitive operational decision. A coordinator responding to interrupted exchanges would require authorized event feeds, timestamps, service guarantees, and latency-aware views. Real-time support remains a conditional research hypothesis, not a capability demonstrated by the figures.

\subsection{Dataspaces for VA: Enabling Conditions and Future Extensions}

\paragraph{Semantic alignment across providers.}
For an analyst comparing institutional catalogs, the Culture Dataspace's information model could supply shared concepts and documented mappings. The scenario illustrates the motivation for that foundation, but its simplified metadata extraction does not demonstrate interoperability across cultural schemas.

\paragraph{Governed enrichment of analytical evidence.}
A collection analyst could extend a coverage analysis with authorized descriptions from additional institutions. The prototype combines fixed public metadata and geometry, illustrating evidence combination rather than dynamic, governed enrichment. An operational extension would require catalog discovery, access negotiation, and reproducible source/version tracking.

\paragraph{Analysis without unrestricted data centralization.}
A coordinator could compare permitted institutional summaries without receiving underlying restricted records. This would require aggregation and disclosure-control mechanisms absent from the public-data demonstrator; the figures do not establish privacy-preserving or federated computation.

\paragraph{Cross-organizational analytical collaboration.}
Curators from different institutions could jointly interpret coverage or quality gaps, but the prototype is a single-user interface. Shared findings, annotations, and permission-aware views would need separate implementation and evaluation, especially where collaborators can see different evidence.

\paragraph{Continuously updated analysis.}
Keeping a metadata coordinator's rankings current would require agreed update interfaces, refresh policies, and explicit treatment of stale or unavailable catalogs. The frozen snapshot demonstrates neither continuous ingestion nor timely updates.

\subsection{Scope of the Demonstration and Limitations}

Together, Figures~\ref{fig:overview}, \ref{fig:bautzen}, and \ref{fig:quality} illustrate complementary design opportunities: understanding federation-wide exchange, inspecting local directional imbalance, and exploring metadata heterogeneity within a shared spatial framework.

\paragraph{Limitations and scope of inference.}
The scenario is semi-synthetic and illustrative: the patterns in Figures~\ref{fig:overview}, \ref{fig:bautzen}, and \ref{fig:quality} must not be interpreted as empirical evidence about real dataspaces or the activities of particular regions or institutions. Real geography and sampled metadata anchor the demonstrator, but regional inventories, quality assignments, and exchange relations are generated. The narrow metadata sample, retained duplicate district identifiers, and heuristic, sample-relative quality scores further limit generalization. The scenario demonstrates how coordinated visual encodings can expose prescribed exchange asymmetries and heterogeneous metadata profiles, and it helps formulate analytical questions and implementation requirements for future monitoring systems. It cannot establish the prevalence or causes of such patterns in practice, absolute metadata completeness, actual provenance or policy compliance, or improvements in governance, efficiency, or user performance. Those claims require representative operational data and empirical evaluation with dataspace stakeholders.

An operational realization would require representative, institution-linked catalogs, validated quality measures, comparable inventory and traffic units, and policy-aware access to exchange events. Record-level lineage would additionally require traceable asset identities and transformation histories, which this prototype does not contain. These requirements illustrate both directions of our vision: visual analytics could make dataspaces more interpretable and actionable, while sovereign dataspaces introduce requirements for visual analytics concerning federation, heterogeneous metadata, and authorized observation across institutional boundaries.

\section{Open Problems and Future Research Directions}

Despite the promising synergy between visual analytics and dataspaces, several fundamental research challenges remain unresolved. These challenges arise from the combination of decentralized data management, heterogeneous data environments, and the need for interactive, human-centered analysis.

\textbf{Bridging decentralized data management and interactive analysis.}  
Dataspaces are inherently decentralized and rely on distributed data sources with varying levels of accessibility and structure. While visual analytics thrives on interactive exploration, enabling such interaction across distributed and potentially restricted data sources remains a key challenge. Existing research on distributed dataspace architectures highlights challenges related to heterogeneity, interoperability, and resource coordination in such environments. Efficient mechanisms are needed to support low-latency, interactive querying and visualization without violating data sovereignty constraints.

\textbf{Scalable visual analytics for heterogeneous and evolving data ecosystems.}  
Dataspaces are characterized by high heterogeneity and continuous evolution of data sources. Visual analytics systems must therefore handle dynamic schemas, incomplete integration, and varying data quality. Existing approaches already struggle with scalability and complexity in large data environments \cite{fekete2012managing}, and these challenges are further amplified in dataspaces where integration is incremental and partial.

\textbf{Semantic interoperability and ontology-driven visualization.}  
Although dataspaces often rely on semantic technologies to establish interoperability, effectively leveraging these semantics in visual analytics remains an open problem. Prior work shows that ontology-based integration is essential for making heterogeneous data analyzable, but translating such structures into intuitive visual representations remains challenging \cite{Ding2020ontology}. Supporting users in understanding semantic relationships across distributed datasets requires new visualization paradigms and interaction techniques.

\textbf{Data provenance, trust, and explainability.}  
Dataspaces emphasize transparency, traceability, and controlled data usage. However, visualizing complex provenance chains and contractual constraints in an understandable way is still unresolved. Provenance data is often large and cognitively complex, making it difficult to represent effectively in visual interfaces \cite{Simmhan2005provenance}. At the same time, explainability remains a major challenge in visual analytics and AI-driven systems.

\textbf{Integration of automated analysis and human-in-the-loop exploration.}  
Visual analytics combines automated analysis with human reasoning, yet balancing these aspects remains difficult. This challenge is intensified in distributed environments such as dataspaces. Research in visual analytics emphasizes the importance of tightly coupling computational methods with human sensemaking, but adaptive workflows that integrate both aspects in decentralized settings remain largely unexplored.

\textbf{Data quality and uncertainty in decentralized environments.}  
Dataspaces often operate under a “best-effort” integration paradigm, where data completeness and consistency are not guaranteed. This introduces significant uncertainty into the analysis process. While uncertainty-aware visual analytics has been studied extensively, effectively communicating uncertainty across multiple data sources and levels of integration remains an open problem \cite{maack2023uncertainty}. This is particularly challenging in collaborative and distributed analysis settings.

\textbf{Usability and accessibility for non-expert users.}  
A central goal of dataspaces is to enable broad participation across organizations. However, current visual analytics tools often require significant expertise. Designing interfaces that allow non-expert users to explore complex, distributed datasets while respecting dataspace constraints remains an open challenge. Research in human–data interaction highlights that usability and cognitive support are still key bottlenecks in visual analytics systems \cite{Keim2010VAprocess}.

\textbf{Standardization and system interoperability.}  
Although architectural frameworks for dataspaces are emerging, the lack of consistent standards for integrating visual analytics components limits interoperability. Existing dataspace research identifies interoperability, semantic alignment, and system integration as key unresolved challenges. There is a need for common abstractions, APIs, and design patterns that allow seamless integration of visualization tools into dataspace infrastructures.

\textbf{Evaluation methodologies for dataspace-enabled visual analytics.}  
Finally, evaluating systems at the intersection of visual analytics and dataspaces poses methodological challenges. Traditional evaluation approaches assume controlled datasets and environments, which contradict the decentralized and evolving nature of dataspaces. More recent research highlights that evaluation in visual analytics must increasingly consider real-world complexity, collaboration, and evolving data contexts. New evaluation frameworks are therefore required.

In summary, the integration of visual analytics and dataspaces opens a rich landscape of research opportunities. Addressing these open problems will require interdisciplinary collaboration across visualization, data management, human–computer interaction, and domain sciences to fully realize the potential of dataspace-enabled analytics.

\section{Vision: Insight-Centric Data Spaces}

We envision a future in which dataspaces and visual analytics are no longer treated as separate layers of the data pipeline, but as tightly integrated components of a unified system. In this vision, dataspaces evolve from passive infrastructures for data exchange into active, human-centered environments for reasoning and decision-making, while visual analytics systems extend beyond isolated tools into distributed, data-aware ecosystems. 
We refer to this convergence as \emph{Insight-Centric Data Spaces} — environments in which dataspaces and visual analytics are tightly integrated to support human reasoning and sensemaking. At the architectural level, we characterize the resulting systems as \emph{Insight-Centric Data Spaces}, emphasizing the active role of human-centered analysis within the data infrastructure.

\subsection{Visual Analytics–Enabled Data Spaces}

In current implementations, dataspaces primarily focus on data exchange, governance, and interoperability. However, they provide only limited support for understanding the data they manage. As dataspaces grow in scale and complexity, this lack of interpretability becomes a critical bottleneck.

Visual analytics can transform dataspaces into interactive environments that support exploration, sensemaking, and decision-making. By embedding visual analytics directly into dataspace infrastructures, users can:
\begin{itemize}
    \item interactively explore distributed and heterogeneous datasets without requiring full integration,
    \item visually analyze data availability, quality, and provenance across participants,
    \item understand contractual relationships, data usage constraints, and access rights,
    \item monitor data flows and detect anomalies or unexpected usage patterns in real time.
\end{itemize}

In this perspective, visual analytics acts as the \emph{cognitive layer} of the dataspace, enabling users to navigate complexity, build trust, and make informed decisions. This is particularly important as dataspaces operate under decentralized control and data sovereignty principles, where transparency and interpretability are essential for adoption.

\subsection{Data Space–Enabled Visual Analytics}

At the same time, dataspaces fundamentally reshape how visual analytics systems are designed and used. Traditional visual analytics assumes centralized, static, and pre-integrated datasets. Dataspaces challenge these assumptions by introducing decentralized, dynamic, and governed data environments.

In a dataspace-enabled setting, visual analytics systems gain access to:
\begin{itemize}
    \item continuously evolving, distributed data sources across organizational boundaries,
    \item semantically enriched data through shared ontologies and metadata standards,
    \item provenance and usage information that supports reproducibility and accountability,
    \item collaborative environments where multiple stakeholders contribute and analyze data jointly.
\end{itemize}

This enables a new generation of visual analytics systems that are inherently distributed, collaborative, and context-aware. Instead of operating on static datasets, these systems interact with live data ecosystems, supporting real-time exploration and decision-making. As data volumes and heterogeneity continue to grow \cite{fekete2012managing}, such capabilities become essential for scalable and meaningful analysis.

\subsection{Towards Insight-Centric Data Spaces}

The true potential lies in the integration of both perspectives. Insight-Centric Data Spaces combine the strengths of dataspaces as decentralized data infrastructures with the strengths of visual analytics as human-centered analysis tools.

In such environments:
\begin{itemize}
    \item data integration becomes an interactive and incremental process guided by analysis needs,
    \item data governance and contracts are made transparent and explorable through visual interfaces,
    \item analytical workflows are distributed across organizational boundaries while remaining interpretable,
    \item users and AI systems collaborate within a shared environment to generate insights.
\end{itemize}

This vision aligns with the broader shift toward human-in-the-loop data analysis, where computational methods and human reasoning are tightly coupled \cite{Thomas2005illuminating}. It also reflects the increasing need to bridge the gap between data management and data understanding in complex, real-world ecosystems.

Ultimately, Insight-Centric Data Spaces represent a step toward fully integrated data ecosystems in which data, analysis, and interaction are no longer separated, but co-evolve within a unified framework. Realizing this vision requires advances in system design, interaction paradigms, and governance models—many of which remain open challenges for future research.

\printbibliography

\begin{IEEEbiography}{Christina Faßbender}
received her Ph.D. in Computer Science from the University of Kaiserslautern in 2018. At the University of Leipzig, she led a subgroup on uncertainty-aware visual analytics for medical applications. She subsequently joined Fraunhofer FIT, where she leads the Data Management and Engineering Group. Her research interests include uncertainty-aware visual analytics, medical data visualization and analysis, and data management.
\end{IEEEbiography}

\begin{IEEEbiography}
{Benjamin Greiner} received his Ph.D. in Physics from TU Dortmund in 2007, focusing on stochastic photon count analysis. At Fraunhofer FIT, he works on digital health, data-driven analytics, and automated analysis of complex scientific data. His research interests include stochastic modeling, data integration, and system architecture. He is a co-inventor of several patents in optical sensing and fluorescence analysis.
\end{IEEEbiography}

\begin{IEEEbiography}{Georgios Toubekis} 
graduated in Architecture at RWTH Aachen University. His research on heritage-led innovation at Fraunhofer FIT  focuses on the use of advanced information systems and the use of data spaces to preserve and valorize heritage. As an expert member of ICOMOS-CIPA Heritage Documentation, he is active in the UNESCO World Heritage Convention framework to protect cultural landscapes worldwide. 
\end{IEEEbiography}

\begin{IEEEbiography}{Benedikt T. Arnold}
received his M.Sc. in Computer Science from the University of Bonn in 2022, where he worked in the Autonomous Intelligent Systems Group. He subsequently joined Fraunhofer FIT, where he researches dataspaces and serves as technical lead for several application projects. His research interests also include large language models and their integration into dataspaces.
\end{IEEEbiography}

\begin{IEEEbiography}{Christoph Lange}
received his Ph.D. in Computer Science from Constructor University (formerly Jacobs University Bremen) in 2011.  He has been working with Fraunhofer since 2014, leading the Data Science and Artificial Intelligence department at Fraunhofer FIT since 2021.  His research, development, consulting and training activities are broadly concerned with knowledge engineering using semantic web technology, and data infrastructures such as dataspaces.
\end{IEEEbiography}

\section*{Acknowledgment}
This work was funded by the Federal Government Commissioner for Culture and the Media, Germany, under grant agreement 2522DIG012, as part of the national flagship lighthouse project “Datenraum Kultur”.

\end{document}